# Preservation of Our Astronomical Heritage
## State of the Profession White Paper for Astro2020
July 9, 2019

Co-Lead Authors:

James Lattis  
Department of Astronomy  
475 N Charter Street  
University of Wisconsin – Madison  
Madison, WI  53706-1507

email:  lattis@astro.wisc.edu

Telephone:  608-263-0360

Wayne Osborn  
Central Michigan University (emeritus)  
118 Eagle Pointe Drive – Unit C  
Delavan, WI  53115

email: Wayne.Osborn@cmich.edu

Telephone:  262-725-6183

Co-authors:

Jennifer Lynn Bartlett, U. S. Naval Observatory (Chair, WGPAH), Washington DC  
Elizabeth Griffin, Herzberg Astronomy & Astrophysics Research Centre, Victoria, Canada  
Thomas Hockey, University of Northern Iowa, Cedar Falls IA  
Stephen McCluskey, West Virginia University, Morgantown WV  
Terry Oswalt, Embry Riddle Aeronautical University, Daytona Beach FL  
Alexei A. Pevtsov, National Solar Observatory, Boulder CO  
Sara Schechner, Harvard University, Cambridge MA  
Virginia Trimble, University of California – Irvine, Irvine CA

# Astro2020 State of the Profession White Paper:
# Astronomy's Archival Materials
July 9, 2019

**Abstract**
We argue that it is essential that the Astro2020 survey of the present state of American astronomy and the recommendations for the next decade address the issue of ensuring preservation of, and making more discoverable and accessible, the field's rich legacy materials. These include both archived observations of scientific value and items of historical importance. Much of this heritage likely will be lost if action is not taken in the next decade. It is proposed that the decadal plan include recommendations on (1) compiling a list of historic sites and development of models for their preservation, (2) carrying out a comprehensive inventory of astronomy's archival material, and (3) digitizing, with web-based publication, those photographs and papers judged to have the most value for scientific and historical investigations. The estimated cost for an example project on plate preservation is a one-time investment of less than $10 million over ten years plus the typical on-going costs to maintain and manage a medium-sized database.

**Key Issue and Overview of Impact on the Field**

There is an urgent need to develop a comprehensive strategic plan, and identify resources, for preserving astronomy's legacy of valuable archival materials. The rich trove of materials can be roughly divided into two main categories. First, there are what we denote as **archival data and observations** – materials of certain or potential scientific value such as the extensive collections of astronomical photographs and their associated metadata as well as drawings, observer's notebooks and other observational records that are data from periods in the time domain that cannot be otherwise accessed. Second, there are what we denote as **historical items** – the papers, letters, and notebooks of astronomers and their organizations as well as physical artifacts such as instruments and observatories that are necessary for understanding how astronomy has developed. Loss of key materials from either category will negatively affect our field.

Astronomical discoveries have changed our understanding of the universe in revolutionary ways. Preserving a representative sample of materials – including instruments – that demonstrates this progress is essential to understanding how we have come to our current view of the universe, how astronomy relates to other sciences, and how the culture of astronomy has evolved. As professionals, we have a responsibility to our predecessors and to our successors to conserve this legacy and ensure that it remains available to future scientists and historians of science.

Over the past two decades, awareness of our multifaceted astronomical heritage, and of threats to it, has been growing. The rapid evolution of the discipline, and especially the closing of famous and historically significant observatories, prompted the AAS to create in 2007 the Working Group on the Preservation of Astronomical Heritage (WGPAH). This working group is charged with "developing and disseminating procedures, criteria and priorities for identifying, designating, and preserving astronomical structures, instruments, and records so that they will continue to be available for astronomical and historical research, for the teaching of astronomy, and for outreach to the general public." [1] In 2018 the International Astronomical Union (IAU) adopted Resolution B3 recommending "that a concerted effort be made to ensure the preservation, digitization, and scientific exploration of all of astronomy's historical data, both analogue and primitive digital, and associated records." [2]  This includes ensuring that what is being collected and used today remains accessible to future astronomers and historians. Therefore, concern for the heritage of our field is incumbent on all practitioners, not just a select few specialists.

*Present Status of the Legacy Materials*
We now survey the current state of the legacy materials, first the archival data and observations and then the historical items.  Legacy materials often have both scientific and historical relevance.  For example, an astronomer's journal may contain scientifically useful observations (*e.g.*, sunspot drawings, brightness estimates of a historic nova) but also provide key historical evidence (*e.g.*, showing unrecognized priority on an important scientific discovery).  A photographic plate may record objects of current astronomical interest but also have annotations on the plate of historical value. In addition, many legacy items, such as unusual instruments or journals of prominent astronomers, elicit great public interest, which makes them valuable tools for "broadening participation" in astronomy, as well as science in general, through education and public outreach (EPO) programs, citizen science activities, and presentations.  The importance of a citizenry that appreciates the importance of science cannot be underestimated for any nation.

1. **The archival data and observations of scientific value**

Given the long time scales of some astronomical phenomena, older astronomical observations are of great value. There is also a growing awareness that the data science techniques developed to support the new large astronomical surveys also provide opportunities to mine older observational archives in ways that their creators could not have foreseen.  Nevertheless, because much of the older data (photographs, drawings, written descriptions) are not easily discoverable and not in digital form, their use usually requires physical access and tedious measurement, severely limiting their use in modern astronomy.  Fortunately, some progress is being made in resolving this problem, such as Harvard Observatory's DASCH plate scanning project [3] and the American Association of Variable Star Observers (AAVSO) International Database (AID) program [4] putting archival variable-star observations on-line.

On the negative side, there are a number of cases where potentially valuable data have been lost, often due to lack of planning for their preservation or for financial considerations. Examples are many, including the discard of early Michigan Curtis Schmidt plates and the Menzel Gap in the Harvard sky patrol series resulting from a budget-reducing suspension of plate taking for several years.  It is time to ask:  Who is responsible for maintaining project data when the principal investigator retires? How can we ensure preservation of valuable data in cases of loss of funding?  When a custodian of archival data (*e.g.*, a traditional observatory) changes its mission and decides to "purge" its holdings, how should decisions be made on what data and historical artifacts are worth preserving?

A less recognized but important problem involves the discoverability of archived materials. Observations, catalogs and digitized copies of data often are kept by the owner, who places them in his/her own archive.  Such archives are commonly not public nor linked with each other. Thus, there may be multiple (but hard-to-find) collections of historical observations of the same object or particular event (*e.g.*, solar eclipse). Standards and methods for discoverability of legacy material require urgent attention. Digitizing a large quantity of historical data without establishing a mechanism for finding and linking it to associated or similar data elsewhere leads to digital "landfills" of very limited use. The related issue of standardization (data formats, required metadata, keywords, *etc*.) also urgently needs to be addressed.

*Recent archival data (primarily of scientific interest):* The more recent archival data are generally well archived and accessible in digital form. The National Aeronautics and Space Administration (NASA) has long had a policy of archiving observations from space missions and space observatories, many *via* the Mikulski archive (MAST). In accordance with Astro2010 recommendations that data from major ground-based programs be similarly archived [5], some organizations have begun doing so. For example, the national observatories automatically archive raw observational data through the "Save the Bits" program. However, some programs still struggle to find a workable archiving strategy; for instance, the High Energy Astrophysics Science Archive Research Center (HEASARC) generally does not ingest results from current high-energy missions borne by sounding rockets.
**Overall status**: very good.
**Recommendation**: Such archiving should continue, and be expanded when possible, with the necessary financial support being built into the budgets of major programs and missions from their initiation.

*Recent published data (primarily scientific):* Astronomical data included in research papers published in the last fifty years are well archived and easily available through the cooperative efforts of the journals, the ADS, arXiv and the CDS in Strasbourg, France (SIMBAD, Vizier). These archiving services are outstanding examples of how scientific data can be preserved and made accessible for future generations and exemplars of what other sciences should be doing. But, in cases where a paper's underlying observations were not published or archived, they can be difficult to obtain. Some journals now request such data be archived with a link to their repository given.
**Status**: very good.
**Recommendation**: Appropriate participation in and support for these international collaborations should be continued and expanded if needed. Authors of research papers should be encouraged to publish or publicly archive the underlying data.

*Early electronic images and records (primarily scientific):* Recording of astronomical observations in digital formats commenced around 1980 and, within ten years, it largely replaced photographic techniques. Unfortunately, the initial generations of digital data were poorly archived, if at all. Unless regularly migrated to new media as data technology improved, these observations remain on media (*e.g.*, punched cards, magnetic tapes, floppy disks, hard drives) that are now inaccessible because either the hardware to read them is no longer available, the format is obsolete, or the underlying medium itself has degraded. Without remediation, these records will be lost. As an example, the earliest X-ray data from balloon- and rocket-borne instruments were recorded on 7-track tapes. Despite the better X-ray observations achieved by the Chandra X-ray Observatory or XMM-Newton, the older observations would be useful in placing constraints on the time evolution of X-ray sources that cannot be attained in any other way. Although the National Space Science Data Center (NSSDC) tracks information about US space missions, including sounding rocket launches, it does not make the actual data available.
**Overall status**: poor to fair.
**Recommendation**: Standards should be developed for the archiving of electronic data (*e.g.*, CCD images, spectra). They should address which types of archives are needed (*e.g.*, for sky patrol images, stellar spectra, solar data, comet observations, planetary imaging, specific space missions), what volume of data should be kept (*e.g.*, all versus a sampling for high-volume surveys), how the data should be archived (perhaps cloud storage so that image files are automatically updated as storage formats evolve), and the entities responsible for managing and underwriting the archives. Whenever possible, we recommend the original media be conserved along with appropriate hardware to read them.

*Archived observations on photographic plates (primarily scientific):* The continuing scientific value of astronomical photographic observations, largely on glass plates, has been established in numerous publications [6]. However, these data are under-utilized because digital copies rarely exist, so it is difficult for a researcher to determine what is available and he/she must physically access the plates to use them. Limited use leads to an erroneous perception of limited value, to neglect, and degradation. WGPAH works to raise awareness of the untapped potential of these collections. Recent examples include two conferences focused on addressing the barriers to exploiting these data [7].

Lack of funds is the single biggest obstacle to increasing access to these collections. An Astro2010 white paper [8] argued that allocating funds for such archival data work is well worth the investment. For example, the digitization of Harvard's photographic plates (DASCH) has been highly productive scientifically. Researchers involved in time-domain studies now routinely use the DASCH database releases, which cover the sky from roughly 1890–1990 with accurate positions and photometry. Concern about plate collections being discarded led to establishment of the Astronomical Photographic Data Archive (APDA) by the Pisgah Astronomical Research Institute (PARI). APDA has the mission of maintaining plate collections that can no longer be housed at their original institutions. APDA now hosts over 50 collections totaling some 340,000 plates [9], and provides researchers digital copies of plates upon request. However, insufficient resources, both financial and manpower, have impeded making this large resource fully accessible.

Most other North American institutions with significant plate collections are doing little to make these resources easily discoverable and accessible. America generally lags behind Europe and China, where coordinated plate digitization projects are occurring (*e.g.*, the German Archives of Photographic Plates for Astronomical Use (APPLAUSE) [10]).

**Overall status**: very mixed.

**Recommendation**: A program should be undertaken to prepare finding aids (catalogs) for major North American plate collections as a precursor to digitizing those plate series of definite scientific value.

## 2. The historical materials

While records of earlier observations have potential to support new scientific research, some legacy items are primarily of historical importance. They provide unique perspective not only on the tremendous strides in astronomy during the 20th century that led to our current understanding of the universe, but also how astronomy relates to other sciences, and how the culture of modern astronomy has developed. Preservation of these items has traditionally not been a high priority, but methods and standards for preserving this legacy, including not only papers and written records but also key instruments and facilities (for example, the major 20th century observatories) before they are lost should be an integral part of the strategic plan for American astronomy in the coming decade.

*Records of individual astronomers (primarily historic)*: Astronomers' personal records of historical value include paper documents such as notebooks, correspondence, diaries, calendars, photographs, working notes, and other items. Materials that have been deposited in an institutional archive or community archive, such as the American Institute for Physics (AIP) Center for the History of Physics, are well preserved and organized. While little of this material is digitized or available on-line, the institutions often have on-line finding aids for their holdings. Along with preserving physical ephemera of an astronomer's professional life, the AAS Historical Astronomy Division (HAD) is capturing the personal experiences of astronomers in their own words through an oral history project [11].
**Overall status**: good.
**Recommendation**: Astronomers and their employers should be reminded that archiving astronomers' papers is important, so that items of potential interest are not discarded when someone retires or dies. A "best practices" guide in identifying "inherited" material worth preservation should be prepared for those who are not professional archivists but find themselves in possession of such items.

*Records of Astronomical Institutions (primarily historic)*: Government agencies, such as the National Science Foundation (NSF) and NASA, generally have a policy on records retention. Therefore, we assume that their records of historical interest indeed are being preserved. Whether these are sufficiently discoverable through finding aids needs further investigation. The overall condition of this material is probably good. We recommend that these institutions also include provisions for "finding aids" and accessibility for records in their retention policies.

Professional organizations, *e.g.*, the AAS, Astronomical Society of the Pacific (ASP), and AAVSO, generally recognize that their papers should be preserved for their historical interest. Some organizations have a person or committee charged with this task. As in the case of governmental agencies, the ability to locate items of interest is not clear.

Record preservation and archiving in institutions whose main mission is research, such as active observatories, are sometimes not priorities. A related problem occurs when a historic observatory closes: What should be done with the institutional papers? Fortunately, in recent cases of closure, the responsible institution has recognized the importance of an observatory's paper records and arranged that they be preserved somewhere.
**Overall status**: fair to good, depending on type of institution.
**Recommendation**: Those institutions without a record retention policy should create one that will ensure that items of historical interest are preserved. Record retention policies should include provisions for cataloging, on-line finding aids, and accessibility.

*Historically important instruments (primarily historic):* Several institutions, *e.g.*, the Adler Planetarium, the National Air and Space Museum, and Harvard University, have assembled collections of, and continue to acquire, instruments of scientific importance. These collections are organized and professionally managed, and their holdings are available for research. It is important that collection policies not neglect significant instruments of the 20$^{th}$ century in favor of those whose only distinction is greater age.

The treatment of historically important instruments at active observatories is more problematic. Outdated or retired instrumentation is typically put in dead storage, often with the result that its history is lost, and it is cannibalized or discarded. Modernization of instruments to prolong their scientific life can remove features of historic value.

When an observatory closes, telescopes or other instruments may be donated to organizations desiring to make further use of them or be scrapped. The uneven handling of such instruments is illustrated through four examples: (1) When David Dunlap Observatory closed, instruments except the 74-inch telescope were relocated to the University of Toronto (with a few donated to the Canada Science and Technology Museum); (2) When Swarthmore College repurposed Sproul Observatory, the 24-inch refractor was donated to a science center, Northwest Arkansas Space (NWA Space) [12], but the status of the other instruments is unclear; (3) The University of Chicago compiled an inventory of Yerkes Observatory equipment while closing the facility, but the eventual disposition of the instruments is uncertain; and (4) the University of Illinois appears to have lost Joel Stebbins' original photoelectric photometer.

Instrumentation belonging to government facilities faces many of the same challenges as those at active observatories, with the additional concern that instruments are more often unique. Items utilized in space missions are, with rare exceptions, unavoidably lost, but their prototypes can be preserved.

**Overall status**: very mixed (from poor to good, depending on institution).

**Recommendations:** Institutions should identify and maintain an on-going inventory of their major instruments. If an institution cannot maintain its instrument collection appropriately, it should consider offering pieces to organizations specializing in astronomical instrument collections or museums with a space-exploration focus. (The WGPAH can assist in making such connections.) Before a historically significant instrument is donated, sold, scrapped or warehoused, a record (including photographs and any engineering documentation) should be appended to the equipment inventory. The astronomical community should recognize the historical value of instrument preservation efforts and provide the modest financial support such work requires.

*Historically important observatories (primarily historic)*: The United Nations Educational, Scientific and Cultural Organization (UNESCO) has designated one United States astronomical observatory (Mount Wilson) as a World Heritage site. The list of National Historic Landmarks names six optical and two radio observatories [13], and the National Register of Historic Places lists thirty-three more [14]. However, most of the more prominent and historic observatories *(e.g.*, Lick, Palomar, and Yerkes) are not on either list, probably because they still are active, or were until recently. As observational astronomy evolves, astronomers must decide what to do with an observatory when it is no longer of value for research. Several historically important observatories have been successfully preserved by transitioning them into museums or outreach centers for astronomy education, including Lowell's Mars Hill Complex and Mount Wilson. On the other extreme, observatories often occupy valuable land and their managing institutions can be tempted to sacrifice the historical value of the facility for redevelopment purposes. Shattuck Observatory at Dartmouth College [15] and Yerkes Observatory of the University of Chicago [16] have weathered such challenges recently. The National Radio Astronomy Observatory (NRAO), Lick, and Palomar are among historically important observatories facing possible future preservation challenges. This issue has stimulated an initiative to create the Alliance of Historic Observatories (AHO, currently in formation) to help such facilities find a path forward. EPO centers, tourism, consortium management, and some development are the parts of models that can be considered to ensure preservation.

**Overall status**: fair.

**Recommendation**: The community should compile a list of significant historic observatories, and study models for their preservation and continued use, where applicable, for educational, museum or other purposes when they are deemed no longer scientifically viable. Institutions desiring to dispose of such sites should pursue a repurposing model that ensures observatory preservation over a future involving demolition or destruction of its character. (The WGPAH stands ready to assist possible stakeholders lacking astronomical expertise.) Funding sources should assist in making the transition to a new purpose possible.

*Archeoastronomical sites (primarily historic)*: Systematic study of the astronomical connotations of ancient and prehistoric sites began in the early 1960s with Stonehenge and other monolithic sites in Britain [16]. By the 1970s such work had spread around the world, and various sites and rock art in the United States were argued to have astronomical significance. One can mention Chaco Canyon, the Bighorn Medicine Wheel, the Cahokia Woodhenge, and the California rock art sites. The importance of preserving such archeological sites of cultural significance now is generally recognized. The IAU has been active in providing a methodological framework for the identification, evaluation, and preservation of heritage sites related to astronomy and archaeoastronomy [17].

Protection and preservation of an archeoastronomical site involves more than an element, say a building or a panel of rock art, with an astronomical connection. It also includes the immediate physical context, the nearby surroundings and associated antiquities, and the viewscape. Sites that involve alignments to the horizon require protection of the horizon and the view. Sites that involve displays of light on architecture or art require the field of projection to be preserved.

A list of ancient sites in the United States with an astronomical dimension would include many dozen candidates. Their governance varies greatly because they are under many different jurisdictions, both governmental (local, county, state, federal, tribal) and private (individual, family, business, corporation, foundation, non-profit). Some are protected, while others are at risk of damage or destruction. The disparate nature of site ownership requires astronomical community involvement, including consultation and work with a wide variety of stakeholders: indigenous people, the archaeological community, private entities, and public agencies. The WGPAH recognizes our responsibility to support the protection and preservation of the U.S. archaeoastronomical heritage and provide astronomical expertise when needed.

**Overall status**: highly variable, depending on case.
**Recommendation**: A committee, under the auspices of the WGPAH, should be established to (1) develop a list of U.S. archaeoastronomical sites using the ICOMOS/IAU Thematic Study [18] criteria, and (2) assist stakeholders in such sites in efforts to preserve them.

**Strategic Plan:**

This coming decade is the time to systematically assess the overall condition of our astronomical heritage and its value from both scientific and historical perspectives given that much of this legacy is in danger of being lost. Priority must be given to preserving those resources of most value for astronomy and at greatest risk and to developing best practices for handling emerging situations such as the coming "huge data" observational programs.

Many recommendations have been presented in this report, and no single plan can address all. Instead, we argue that the decadal strategic plan should explicitly recognize the importance of preservation of astronomical heritage. Some of our views might be included as part of the recommendations for proposed space missions, ground-based programs, and other activities.

**Organization, Partnerships, and Current Status**
Concern for our shared astronomical heritage has grown over the past two decades, and many organizations are beginning to engage in preservation efforts. The entire astronomical community shares responsibility for protecting this legacy. Unresolved questions regarding handling of materials, neglect, deferred maintenance, and other factors have put many resources at risk. Coordination among stakeholders will be essential to effectively using the limited resources likely available for preservation and curation, and to press the case for funding to make items more available. Besides the WGPAH, the following organizations have a close connection to community efforts to preserve astronomy's legacy: the IAU, HAD (which we understand is submitting a similar "State of the Profession" white paper giving their perspective), the nascent Alliance of Historic Observatories, and several universities, observatories, museums, and other institutions. For its part, the WGPAH commits to working for, and assisting stakeholders in, this important activity.

Our survey of the current status of different types of astronomy's legacy materials leads to three general priorities for action in the next decade. First, the community should compile a list of sites valued for their historic significance and explore models for their preservation and development, for example as educational and museum facilities, or other appropriate uses. Second, the community should collaborate in carrying out a comprehensive inventory of archival materials, which would include archival data, early electronic images and records, the records and papers of astronomers, their instruments, and institutions. Third, the community needs to judge which of its papers and photographic data have high value for scientific and historical investigations and undertake efforts to make digital copies available through the web.

**A Preservation Project Example: Astronomical Plate Data**
A heritage preservation project involving the astronomical plate collections offers a tractable example of how a modest investment could yield significant scientific dividends. The Harvard DASCH program has demonstrated the value of digitizing these past records of the sky. It is estimated that North American institutions hold about 1.3 million direct photographic plates plus about one million spectrograms, solar images, and other photographic records [19]. Over 95% are in just thirteen institutions. A digitization program is needed to make this trove of data accessible for modern scientific utilization. But not all photographic records are worth digitizing, so the first step should be an inventory of the major plate collections followed by the development of finding aids (on-line catalogues) and priorities for scanning.

An inventory of the major plate collections, the development of finding aids for sky fields, objects and spectra, and high-quality digitization of selected plate sets to complement the Harvard ones or of unique series, likely could be accomplished in ten years. A dedicated group, such as WGPAH, could manage such a program, although local institutions could be as effective, as in the Harvard-DASCH model. The projected cost is under $10 million. To date, however, requests for support of plate cataloguing and digitization (typically in the <$200 K range) have usually been rejected, both by agencies that fund science and those that fund preservation of historical materials. What is essential, regardless of the final form any preservation project takes, is that funds be made available. A possible timeline and a cost estimate for our example project follow.

**Schedule (for plate preservation project example):**
A time line for a project to produce catalogs of major plate archives and the digitization of selected plates of scientific value (both direct images and spectra) is:

Year 1         Secure the funding commitment and organize the process.
Years 2 – 4    Call for proposals and award of funds to institutions with significant plate collections for carrying out an inventory of their plates and creating catalogs.
Years 5 – 6    Select those plate sets most worthy of digitization. Coordinate with DASCH (or elsewhere) on use of their scanner, or construct DASCH clones or other reliable high-resolution scanners. Commissioning tests of scanning systems.
Years 7 – 10   Fund scanning projects for the selected plate sets.

**Cost Estimates (for plate preservation project example):**
Based on work with photographic archives over the past ten years, both digitization projects such as DASCH and inventory and cataloging projects such as done at Yerkes and Carnegie Observatories, it is estimated that on-line catalogs of the major plate collections could be prepared in three years. The digitization of those items given high priority could be completed in ten years utilizing two to three high-speed, high-resolution custom-built scanners such as the DASCH machine. Experience points to a ballpark total cost of under $10 million over a ten-year period, with costs spread over several institutions. Labor costs would be the most significant fraction of a budget. This would be a one-time investment. Subsequent on-going costs (database maintenance, web interface, etc.) would be similar to the operating costs of any medium-sized digital database.

**Conclusion:**
The acquisition, preservation, accessibility, and distribution of historically important documents, data, instruments, and facilities are an ongoing responsibility of the astronomical community that requires continual and stable funding through the coming decade and beyond. That said, an urgent need in the coming decade is to carry out comprehensive inventories, where lacking, of those legacy materials, particularly the plates, the historical observatories, and other astronomical sites.

We strongly recommend that the preservation and archiving activities outlined above should be endorsed by the ASTRO2020 Strategic Plan. This, of course, implies financial commitment from the astronomical community and its funding agencies. That could be obtained *via* funding specific preservation and archiving projects (such as a plate cataloging and digitization effort) or by requiring that data preservation and archiving be included in the budgets of new and renewal proposals. A combination of these two modes of funding would be best. Along similar lines, plans for the preservation of astronomically important artifacts, instrumentation, and facilities should be a required part of research funding requests and construction projects.

**References**

[1] AAS. 2007, "Charge of the Working Group on the Preservation of Astronomical Heritage WGPAH)"
https://aas.org/comms/working-group-preservation-astronomical-heritage-wgpah

[2] Lago, T. 2019, *Transactions IAU*, Vol. XXXB, Resolution B3.
https://www.iau.org/static/archives/announcements/pdf/ann18029d.pdf

[3] Grindlay, J. 2019, "Digital Access to a Sky Century @ Harvard (DASCH)."
http://dasch.rc.fas.harvard.edu/project.php

Grindlay, J. E. 2014, AAS Meeting #224, paper id. 118.01.
http://adsabs.harvard.edu/abs/2014AAS...22411801G

[4] AAVSO. 2017, AAVSO International Database.
https://www.aavso.org/aavso-international-database

[5] *Committee for a Decadal Survey of Astronomy and Astrophysics*. 2010, New Worlds, New Horizons (Washington: National Academies Press), p. 146.

[6] Examples are:
Hippke, M., *et al.* 2017, *ApJ*, 837, 85.
https://iopscience.iop.org/article/10.3847/1538-4357/aa615d/pdf

Schaefer, B. E. 2016, *ApJ*, 822, L34.
https://ui.adsabs.harvard.edu/abs/2016ApJ...822L..34S/abstract

Osborn, H. P., *et al*. 2019, *MNRAS*, 485, 1614.
https://academic.oup.com/mnras/article-abstract/485/2/1614/5304181?redirectedFrom=PDF

For an earlier, more complete, compilation see Osborn, W. & Robbins, L., 2009, *ASP Conference Series*, 410, ed. W. Osborn & L. Robbins, Appendix D, p. 196.
http://articles.adsabs.harvard.edu/pdf/2009ASPC..410..196.

[7] Osborn, W. & Robbins, L. 2009, *ASP Conference Series*, 410, ed. W. Osborn & L. Robbins, p. 33. http://articles.adsabs.harvard.edu/pdf/2009ASPC..410...33O

Osborn, W. & Lattis, J. 2013, *PASP*, 125, 113.
https://iopscience.iop.org/article/10.1086/669131/pdf

[8] Osborn, W., Accomazzi, A., Castelaz, M. et al. 2009, *ASP Conference Series*, 410, ed. W. Osborn & L. Robbins, Appendix A, p. 160.
https://ui.adsabs.harvard.edu/abs/2009ASPC..410..160O/abstract

[9] Barker, T. et al., 2019, AAS Meeting #233, paper id. 262.04.
https://ui.adsabs.harvard.edu/abs/2019AAS...23326204B/abstract

[10] APPLAUSE. 2019, "Archives of Photographic PLates for Astronomical USE," *APPLAUSE* (Deutsche Forschungsgemeinschaft).
https://www.plate-archive.org/applause/

Groote, D., Tuvikene, T, Edelmann, H. *et al.* 2019, *Astroplate 2014, Proceedings of a conference held in March, 2014 in Prague, Czech Republic*, ed. L. Mišková & V. Stanislav Vítek (Prague: Institute of Chemical Technology), p. 53.

[11] HAD. 2019, "Astro2020 State of the Profession White Paper: History of Astronomy," in preparation.

Buxner, S., Holbrook, J. & AAS Oral History Team. 2016, AAS Meeting #228, id. 214.13. https://ui.adsabs.harvard.edu/abs/2016AAS...22821413B/abstract

[12] Eley, A. 2017, "Large telescope moves to Northwest Arkansas to further STEM recruitment goals," *Northwest Arkansas Democrat Gazette* (Fayetteville, AR: Northwest Arkansas Newspapers) (retrieved 2019 July 1). https://www.nwaonline.com/news/2017/jul/30/large-telescope-moves-to-northwest-arka/

[13] NPS. 2019, "National Register Database and Research," *National Register of Historic Places* (Washington, DC: U.S. Dept. of Interior) (retrieved 2019 June 26). https://www.nps.gov/subjects/nationalregister/database-research.htm

[14] NPS, 2019. *List of National Historic Landmarks* (Washington, DC: U.S. Dept. of Interior) (retrieved 2019 July 4). https://www.nps.gov/subjects/nationalhistoriclandmarks/list-of-nhls-by-state.htm

[15] Jones, C. 2018, Letter to Dartmouth College regarding Shattuck Observatory. https://aas.org/files/aas_president_letter_to_dartmouth_president_re_shattuck_observatory_feb2018_0.pdf

*The Valley News* (Lebanon, New Hampshire), 2017 Nov. 12 (retrieved 2019 July 4) https://www.vnews.com/Shattuck-Observatory-Astronomy-Department-13619574

[16] Donahue, M. 2018, Letter to University of Chicago regarding Yerkes Observatory. https://aas.org/files/aas-donahue-yerkes-letter.pdf

For more information see: https://en.wikipedia.org/wiki/Yerkes_Observatory

[17] Hawkins, G. S. & White, J. B. 1965, *Stonehenge Decoded* (Garden City Nee York: Doubleday and Co.).

[18] Ruggles, C. and Cotte, M., ed. 2011, *Heritage Sites of Astronomy and Archaeoastronomy in the context of the UNESCO World Heritage Convention: A Thematic Study* (Paris: International Council on Monuments and Sites / International Astronomical Union)

[19] Osborn, W. & Robbins, L. 2009, *ASP Conference Series*, 410, ed. W. Osborn & L. Robbins, p. 81. http://articles.adsabs.harvard.edu/full/2009ASPC..410...81R